\documentclass[12pt]{iopart}

\usepackage[pdftex]{graphicx}
\usepackage[pdftex,bookmarks=true,bookmarksopen,bookmarksnumbered,
                colorlinks,
                linkcolor=blue,
                citecolor=blue]{hyperref}
\usepackage{dcolumn}
\usepackage{bm}
\usepackage{multirow}

\begin{document}

\title{Broadband Purcell enhanced emission dynamics of quantum dots in linear photonic crystal waveguides}

\author{A Laucht$^{1,2}$, T G\"unthner$^{1,3}$, S P\"utz$^1$, R Saive$^1$, S Fr\'ed\'erick$^{1,4}$,\\N Hauke$^1$, M Bichler$^1$, M-C Amann$^1$, A W Holleitner$^1$,\\M Kaniber$^1$, and J J Finley$^1$}

\address{$^1$ Walter Schottky Institut and Physik-Department, Technische Universit\"at M\"unchen, Am Coulombwall 4a, 85748 Garching, Germany\\
$^2$Centre for Quantum Computation \& Communication Technology, The University of New South Wales, Sydney NSW 2052, Australia\\
$^3$Institut f\"ur Experimentalphysik, Universit\"at Innsbruck, Technikerstrasse 25, 6020 Innsbruck, Austria\\
$^4$National Research Council of Canada, Ottawa, ON, Canada\\}
\ead{finley@wsi.tum.de}

\begin{abstract}
The authors investigate the spontaneous emission dynamics of self-assembled InGaAs quantum dots embedded in GaAs photonic crystal waveguides. For an ensemble of dots coupled to guided modes in the waveguide we report spatially, spectrally, and time-resolved photoluminescence measurements, detecting normal to the plane of the photonic crystal. For quantum dots emitting in resonance with the waveguide mode, a $\sim21\times$ enhancement of photoluminescence intensity is observed as compared to dots in the unprocessed region of the wafer. This enhancement can be traced back to the Purcell enhanced emission of quantum dots into leaky and guided modes of the waveguide with moderate Purcell factors up to $\sim4\times$. Emission into guided modes is shown to be efficiently scattered out of the waveguide within a few microns, 
contributing to the out-of-plane emission and allowing the use of photonic crystal waveguides as broadband, efficiency-enhancing structures for surface-emitting diodes or single photon sources.

\end{abstract}

\maketitle

Photonic crystal waveguides (PCWs) are of strong interest as optical elements for integrated nanophotonic optical circuits and on-chip quantum optics applications.~\cite{Hughes07, Englund07b, Volkov07, Faraon08b, Yao10b} They have been used to route single photons from cavity-coupled~\cite{Englund07b, Yao09, Englund10, Yao10b} and waveguide-coupled quantum dots (QDs)~\cite{Schwagmann11, Laucht12}, but also to tailor the local density of optical states (LDOS) an emitter experiences. This method to modify the LDOS experienced by an emitter provides a route to engineer the rate and directionality of spontaneous emission.~\cite{Hughes04, Viasnoff05, Lecamp07, Rao07, Rao07b, Stumpf07, Rao08, Lund-Hansen08, Patterson09, Dewhurst10, Sapienza10, Thyrrestrup10, Yao10b, hoang12} This is a key concept to enhance the efficiency of nanoscale light sources such as single photon sources~\cite{Purcell46, Pelton02, Santori02, Claudon10, Fujiwara11, Davanco11} and nanoscale lasers.~\cite{Altug06} Recently, Kaniber et al.~\cite{Kaniber08a} demonstrated a $\sim16\times$ enhanced extraction efficiency for single QDs emitting into the two-dimensional photonic bandgap of a photonic crystal. The photonic bandgap inhibits photon emission into the in-plane direction and redistributes it into out-of-plane modes, effectively increasing the extraction efficiency. However, this comes at the cost of long radiative lifetimes, imposing an inherent jitter in the photon emission time, a source of quantum distinguishability.~\cite{Kiraz04, Chang06} In contrast, strong enhancements of spontaneous emission rates have been observed for QDs in low mode volume, high-Q cavities.~\cite{Englund05, Kaniber07, Englund10} However, these systems require a sophisticated electro-~\cite{Laucht09} or thermo-optical~\cite{Englund07} tuning method to spectrally bring the emitter and cavity mode into resonance.

In this paper we demonstrate the advantage of using a PCW mode to enhance the emission rate and the extraction efficiency of an ensemble of QDs over a wide spectral range of $\sim 18$~meV in the out-of-plane direction. We compare the emission properties normal to the sample surface for QDs in the bulk GaAs, the photonic crystal membrane (PC) and the PCW region. Measurements were made for QDs in and out of spectral resonance with the guided PCW mode. These measurements show a strong enhancement of the photoluminescence (PL) intensity up to $\sim21\times$ for QDs spatially and spectrally coupled to the waveguide mode compared to QDs in the unprocessed bulk material. We attribute this enhancement to a combination of angular redistribution of emission into leaky PCW modes and Purcell enhanced emission into the guided PCW mode with subsequent scattering into leaky modes.

The sample investigated was grown by molecular beam epitaxy and consists of a $500$~nm thick Al$_{0.8}$Ga$_{0.2}$As sacrificial layer, and a $180$~nm thick undoped GaAs layer with a single layer of nominally In$_{0.5}$Ga$_{0.5}$As QDs at its midpoint. The sample has a dot density of $\rho_{QD}>50$~$\mu$m$^{-2}$. A two-dimensional PC formed by defining a triangular array of air holes was realized using a combination of electron-beam lithography and reactive ion etching. PCWs were established by introducing line defects consisting of a single missing row of holes (W1 waveguide). Free standing GaAs membranes were created in a final wet etching step using hydrofluoric acid.

For optical characterization the sample was mounted in a liquid He-flow cryostat and cooled to $10-15$~K. For excitation we used a pulsed Ti-Sapphire laser ($80$~MHz repetition frequency, $6$~ps pulse duration) tuned to the low energy absorption edge of the bulk GaAs ($\lambda_{laser}=815$~nm). Excitation of the QDs and detection of the emitted PL signal was done perpendicular to the sample surface using a $100\times$ microscope objective (NA=0.50). The full-width-half-maximum ($FWHM$) of the excitation laser spot on the sample was determined to be $FWHM\sim1.3$~$\mu$m, while the size of the detection spot had a diameter of $FWHM\sim6.0$~$\mu$m. The QD PL was spectrally analyzed using a $0.5$~m imaging monochromator and detected using a Si-based, liquid nitrogen cooled CCD detector. For time-resolved spectroscopy we used a Si-based avalanche photodiode connected to the side-exit of our monochromator providing a temporal resolution of $\sim350$~ps.

\begin{figure}[t!]
\centering
\includegraphics[width=0.6\columnwidth]{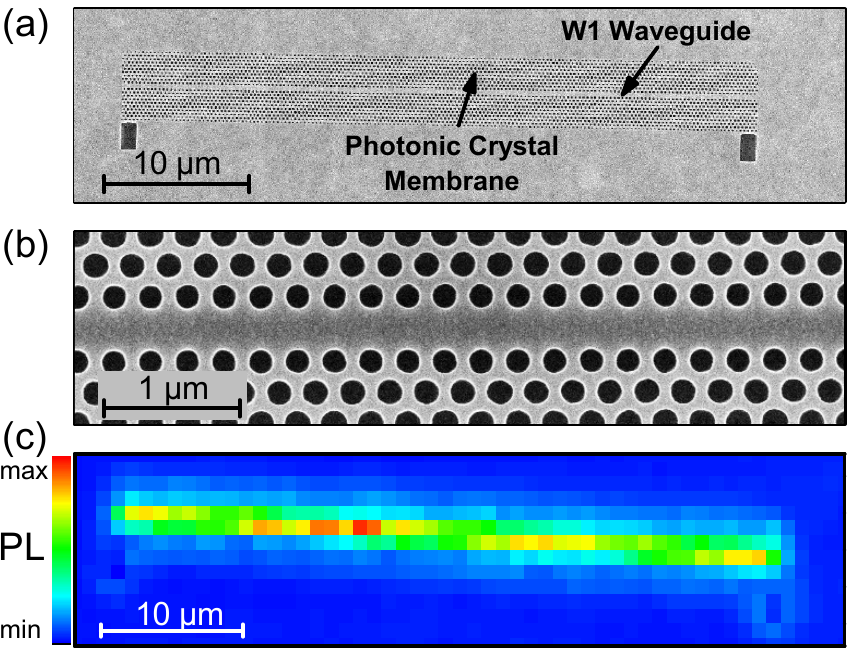}
\caption{\label{figure01} (a) and (b) Scanning electron microscope images of the investigated W1 waveguide - quantum dot system from the top. (c) Spatially-resolved scan of the photoluminescence signal performed with excitation and detection from the top and integrated over the $1298$~-~$1340$~meV spectral range. The area of high photoluminescence intensity corresponds to the waveguide region.}
\end{figure}

In Fig.~\ref{figure01}(a) and (b) we present a scanning electron microscope image of a nominally identical sample to the one used for the PL measurements. The photonic structure has a total length of 45~$\mu$m, a slab thickness of $h=180$~nm, a PC lattice constant of $a=270$~nm, and an air hole radius to lattice constant ratio of $r/a=0.34$. The two rectangles directly next to the photonic crystal structure serve for orientation purposes during optical measurements and have no influence on the photonic properties of the waveguide. We obtained a spatially and spectrally resolved PL map of this particular structure by scanning the laser spot over the surface of the sample and recording a PL spectrum at every position in a confocal detection geometry. A typical result of this scan is presented in Fig.~\ref{figure01}(c) where we have integrated the detected PL over the spectral range of the QDs ($1298$~meV to $1340$~meV) and plotted the resulting intensity in a false color representation. The contours of the photonic structure can be easily recognized in the PL map and we observe a clear luminescence enhancement on the photonic structure compared to the unpatterned substrate.

\begin{figure}[t!]
\centering
\includegraphics[width=0.52\columnwidth]{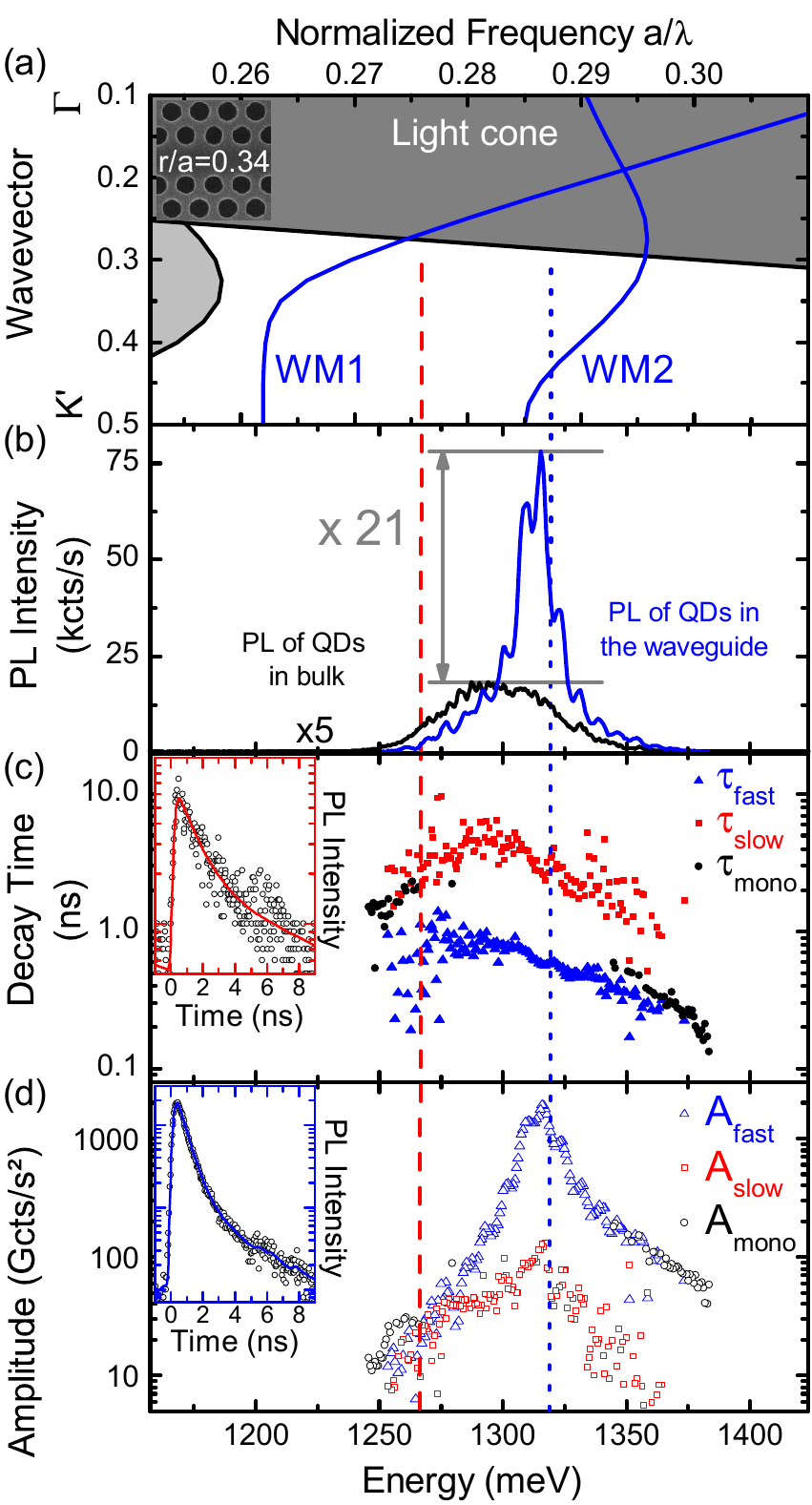}
\caption{\label{figure02} (a) Photonic bandstructure calculations for a W1 waveguide with $r/a=0.34$ and $h/a=0.6667$. The blue lines correspond to the photonic waveguide modes and the light gray region to the slab waveguide modes. The dark gray region indicates the region above the light cone. (inset) Scanning electron microscope image of the investigated photonic crystal waveguide structure. (b) Comparison of the photoluminescence signal measured on the ensemble of quantum dots in the unprocessed bulk material (black line) and on the ensemble of quantum dots in the waveguide region (blue line). (c) and (d) Fitting parameters of the decay transients measured on the photonic crystal waveguide as a function of energy. In (c) the extracted decay times and in (d) the corresponding amplitudes are plotted. The black circles correspond to the decay parameters obtained from mono-exponential fits, while the blue triangles (red squares) correspond to the fast (slow) component of the bi-exponentially fitted decay transients. (insets) Decay transients and fits at the energy of 1275 meV (indicated by the red dashed line) for the upper, red-framed inset and 1318 meV (indicated by the blue dotted line) for the lower, blue-framed inset.}
\end{figure}

We now investigate the origin of this PL enhancement. To do this, we performed photonic bandstructure calculations using the software package RSoft.~\cite{rsoft} We use the appropriate parameters for GaAs of $n_{GaAs}=3.5$ and the geometric parameters of $h/a=0.6667$ and $r/a=0.34$, corresponding to the investigated W1 PCW. The result of this simulation is plotted in Fig.~\ref{figure02}(a) where we plot the normalized frequency of the photonic bands as a function of k-vector on the path from the $\Gamma$ point to the $K'$ point.~\cite{Johnson00,Dorfner08} We present the PCW modes as blue solid lines, the slab waveguide modes as light gray regions, and the lossy region above the light cone is shaded in dark gray. The region above the light cone corresponds to the energy-wavevector combinations for which photons are not confined to the slab by total internal reflection, i.e. they can leave the waveguide in vertical directions. We calculate the guided part of the lowest energy waveguide (0th order) mode WM1 to be at a normalized frequency of $a/\lambda=0.262-0.274$ which corresponds to an energy of $E=1203-1259$~meV. For small wavevectors $k<0.28$ this mode overlaps with the region above the light cone. Photons of these wavevector-energy combinations can, therefore, couple to free space modes and leave the sample in vertical directions. This unguided part of the waveguide mode spans the normalized frequency range of $a/\lambda=0.275-0.308$ ($E=1260-1413$~meV). The second waveguide (1st order) mode WM2 extends from $a/\lambda=0.285-0.296$ ($E=1309-1359$~meV) with an additional unguided region at $a/\lambda=0.291-0.296$ ($E=1338-1359$~meV). The third waveguide (second order) mode WM3 is at much higher energies of $a/\lambda=0.334-0.337$ ($E=1534-1548$~meV) (data not shown). 

For comparison we plot two examples of PL spectra in Fig.~\ref{figure02}(b). The blue line corresponds to a spectrum recorded directly on the PCW, while the black line corresponds to a spectrum recorded next to the PCW on the unprocessed GaAs bulk material (note the $\times 5$ magnified scale). We clearly observe a strong enhancement of the PL intensity in the spectral range between $1300$~meV and $1325$~meV. For these excitation conditions, we detect the maximum signal of $\sim78000$~cts/s at an energy of $\sim1317$~meV, $\sim21\times$ stronger than the $\sim3700$~cts/s obtained from the unprocessed bulk material. The intensity oscillations in the spectrum recorded on the PCW structure, most likely originate from Fabry-Perot resonances due to the finite length of the waveguide.~\cite{hoang12} We notice that the energy range over which we observe PL enhancement coincides very well with the flat part of the dispersion relation of the second energy mode, but also with the energy range of the unguided parts of the first and the second waveguide mode as shown in Fig.~\ref{figure02}(a).

We continue investigating this system by performing time-resolved measurements on the W1 waveguide for 200 different energies in the range between $1239.8\,$meV and $1377.6\,$meV. To do this we used the spectrometer as a spectral bandpass filter of width $\sim 0.25\,$meV and recorded multiple decay transients as a function of energy. The measured data was then fitted with an automated fitting algorithm, taking into account the instrument response function of the experimental setup. The algorithm first tries to fit a bi-exponential decay but automatically switches to a mono-exponential decay when one of the fitting parameters is regarded unrealistic. This is the case when (i) one of the amplitudes becomes negative, (ii) the two lifetimes differ by less than $20\%$, (iii) one of the lifetimes is shorter than $50$~ps, or (iv) one of the amplitudes is more than $25\times$ larger than the other one. In Fig.~\ref{figure02}(c) the extracted decay times are shown and in Fig.~\ref{figure02}(d) the corresponding amplitudes are plotted. The blue triangles (red squares) correspond to the fast (slow) component of the bi-exponentially fitted decay parameters. When a decay is fitted mono-exponentially the resulting parameters are plotted as black circles.

The decay dynamics can predominantly be fitted with bi-exponential parameters close to the resonance with the maximum PL enhancement at $\sim 1317\,$meV, and with mono-exponential parameters off resonance. We interpret this observation according to the fact that only the emission dynamics of QDs spatially located in the waveguide region and spectrally in resonance with the PCW mode can be enhanced by the Purcell effect. However, our experiment detects contributions of spatially coupled \emph{and} uncoupled QDs since our excitation spot size is $4-5\times$ larger than the PCW width. In resonance with the PCW mode the decay times typically exhibit values of $\tau_{fast} = 0.57 \pm 0.1\,$ns and $\tau_{slow}=2.7 \pm 0.1\,$ns. 
The Purcell enhanced decay times are longer than the ones that have been observed for QDs coupled to a photonic crystal cavity mode.~\cite{Englund05,Hennessy07, Kaniber07,Laucht09b} We attribute this to the combined influence of an averaging effect over QDs that are spatially well and badly coupled to the PCW mode, the slope of the second PCW mode in k-space which leads to a lower photonic density of states, and additional broadening of the mode due to fabrication imperfections.\cite{Rao07} 
We also notice a general energy dependence of the decay times. At lower energies we observe longer decay times than at higher energies. We relate this to the enhanced contribution of fast multi-excitonic states to the PL signal at higher energies.~\cite{Adler96, Laucht11} The amplitude of the fast decay component, plotted in Fig.~\ref{figure02}(d), nicely resembles the profile of the enhanced PL spectrum in Fig.~\ref{figure02}(b), while the amplitude of the slow decay component stays almost constant (note the logarithmic scale).

We present two decay transients with fits as examples in the insets of Fig.~\ref{figure02}~(c) and (d). The decay transient in Fig.~\ref{figure02}(c) - inset was recorded off resonance at the energy of $1275$~meV (indicated by the red dashed line), and the transient in Fig.~\ref{figure02}(d) - inset in resonance at the energy of $1318$~meV (indicated by the blue dotted line). Both transients exhibit a bi-exponential decay, albeit with a larger amplitude and shorter decay time of the fast decay component of the resonant transient. This is in good agreement with the extracted parameters presented in Fig.\ref{figure02}~(c) and (d).

At first sight, the observation of a larger emission amplitude in the direction normal to the photonic crystal membrane may seem counter-intuitive. In ideal structures we would expect the Purcell enhanced emission to be efficiently guided away from the excitation spot along the waveguide, resulting in a decrease of emission detected in the vertical direction. However, disorder is known to lead to a scattering of the guided light via Anderson localization and radiation normal to the surface of the waveguide.~\cite{Anderson58,Lagendijk09,Sapienza10} Indeed, recent near field measurements of the frequency dependence of the PCW mode in similar systems have revealed direct evidence for localization and its increasing importance for slow light modes.~\cite{Huisman12} Scattering in the in-plane direction is inhibited due to the photonic bandgap of the surrounding photonic crystal, but scattering into modes above the light cone that can readily escape from the slab is still possible.~\cite{Kuramochi05,Patterson09b,Patterson10} Furthermore, the finite length of the PCW structure also enables the QDs to emit in vertical directions due to Fabry-Perot effects.~\cite{Fussell08} 
In our structures, probably the most important reason for emission in vertical directions is the presence of non-guided waveguide modes with the same energy that can be accessed by scattering induced by disorder. QDs can emit light directly into the leaky modes, or propagating photons from the guided mode can be scattered into these modes due to disorder in the crystal.~\cite{Kuramochi05,Stumpf07,Patterson09b,Patterson10,hoang12} Similar experimental observations have been made been made by Stumpf et al.~\cite{Stumpf07}, and recently by Hoang et al.~\cite{hoang12} and Huisman et al.~\cite{Huisman12}

\begin{figure}[t!]
\centering
\includegraphics[width=0.7\columnwidth]{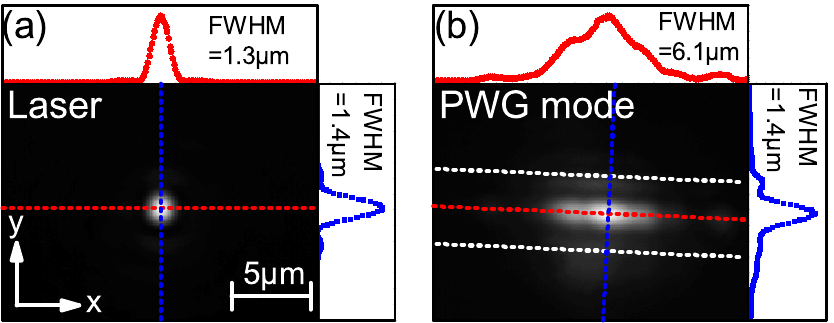}
\caption{\label{figure03} (a) Spatial image of the reflection of the laser spot on the sample surface. (b) Spatial image of the photoluminescence signal of the waveguide mode normal to the sample surface (filtered with a 10nm bandpass filter at the energy of the mode). The dotted white lines indicate the outline of the photonic crystal membrane.}
\end{figure}

In order to support these suggestions we performed additional measurements with a Si-based, Peltier-cooled CCD camera for spatial imaging. In Fig.~\ref{figure03}(a) we present an image taken of the laser excitation spot on the sample surface at an energy of $1521$~meV. The red (blue) curve corresponds to the cross section along the x- (y-) direction at the position indicated by the red (blue) line. From these plots, we can extract the $FWHM$ of the laser spot to be $\sim1.3$~$\mu$m in both the x- and the y-directions. Fig.~\ref{figure03}(b) shows a similar image of the PL emitted by the PCW mode in the vertical direction. In order to record this image, we mounted a $10$~nm-bandpass filter, centered at the energy of the strongest PL enhancement, in front of the camera. While the PL signal recorded along the y-axis, perpendicular to the waveguide direction, exhibits a similar width as the reflection of the laser spot $FWHM=1.4$~$\mu$m, the signal along the waveguide direction (x-axis) is much broader with $FWHM=6.1$~$\mu$m. This observation strongly supports the assertion, that light efficiently emitted into the nominally guided waveguide mode is scattered into vertical directions due to the processes mentioned above. Since our detection spot has a size of $FWHM\sim6.0$~$\mu$m, we collect most of this scattered light and observe, therefore, not only PL emitted into the vertical direction, but also PL emitted into the guided modes and subsequently scattered into the vertical direction.

\begin{figure*}[t!]
\centering
\includegraphics[width=\columnwidth]{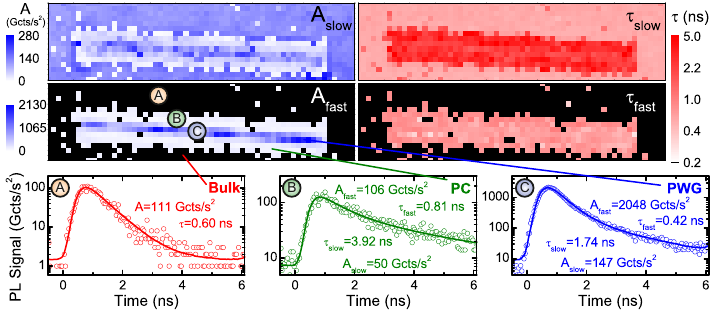}
\caption{\label{figure04} Spatially-resolved scan of the time-resolved photoluminescence signal. The different panels show the fitted amplitude $A_{slow}$ and decay time $\tau_{slow}$ of the slow decay component and the amplitude $A_{fast}$ and decay time $\tau_{fast}$ of the fast decay component. \protect\includegraphics[height=3.7mm]{A}, \protect\includegraphics[height=3.7mm]{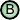} and \protect\includegraphics[height=3.7mm]{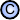} present examples of decay transients of the spatially-resolved scan recorded at the position \protect\includegraphics[height=3.7mm]{A} the bulk material, \protect\includegraphics[height=3.7mm]{B} the photonic crystal membrane, and \protect\includegraphics[height=3.7mm]{C} the photonic crystal waveguide.}
\end{figure*}

We check for Purcell-enhanced emission by performing a spatially-resolved scan of the time-resolved PL signal emitted at the peak of the amplitude ($1317$~meV). In order to do this we scan the excitation spot over the surface of the sample in $1$~$\mu$m-steps and record the decay transients at every position over a 61x16 matrix. The decay transients are then fitted taking into account the instrument response function of the experimental setup. The same fitting algorithm as before was used, which first tries to fit a bi-exponential decay but automatically switches to a mono-exponential decay when one of the fitting parameters becomes unrealistic. The result of the fitting routine is summarized in the different panels of Fig.~\ref{figure04} that show the amplitude $A_{slow}$ and decay time $\tau_{slow}$ of the slow decay component and $A_{fast}$ and $\tau_{fast}$ of the fast decay component. When the decay transient is fitted mono-exponentially, the extracted amplitude and decay time are plotted in the panels for the slow decay component and the corresponding pixel in the panels for the fast decay component are blackened. 

We can clearly recognize the outline of the PCW structure from the data in Fig.~\ref{figure04}. On the unprocessed bulk material the decay transients are globally well described by mono-exponential decays, leading to the large blackened area around the photonic crystal membrane structure (see e.g. position \includegraphics[height=3.7mm]{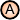} on Fig.~\ref{figure04}). We present a typical example of one of these decay transients in the panel of Fig.~\ref{figure04}, marked by \includegraphics[height=3.7mm]{A}. The amplitude and the decay time of the mono-exponential decay are fitted to be $A_{bulk}=111\pm14$~Gcts/s$^2$ and $\tau_{bulk}=0.6\pm0.1$~ns, which results in an intensity of $I_{bulk}=A_{bulk}\cdot\tau_{bulk}=67\pm14$~cts/s. The extracted parameters of the decays at the different positions are summarized in Tab.~\ref{tabledecays}.

Upon moving away from the bulk material onto the photonic crystal membrane, but not on the W1 waveguide, the decay transients become clearly bi-exponential. A representative transient is presented in Fig.~\ref{figure04} and marked by \includegraphics[height=3.7mm]{B}. Here, we observe a very slow decay component due to the effect of the photonic bandgap and an additional faster decay component probably due to QDs located in proximity to the etched holes of the photonic crystal. The extracted parameters are summarized in Tab.~\ref{tabledecays}. We observe an enhancement in the total intensity of $\sim4\times$ compared to the intensity of the QD ensemble in the bulk material, which we relate to a more efficient collection of the emitted photons due to angular redistribution of emission.~\cite{Fujita05, Kaniber08a} If we take into account the air fill factor of $42$~$\%$ of the photonic crystal structure, i.e. $42$~$\%$ less QDs, the emission efficiency is even a factor $\sim7\times$ higher than that observed from the bulk material.

Next, we focus on the waveguide region. Here, the oscillations in $A_{fast}$ in Fig.~\ref{figure04} are an artifact of the slightly rotated waveguide structure and the 1~$\mu$m steps during the scan. We observe a highly enhanced amplitude for both the fast and slow decay components, as can be seen in the example transient in Fig.~\ref{figure04} marked by \includegraphics[height=3.7mm]{C}. The extracted parameters are summarized in Tab.~\ref{tabledecays}. With a total intensity of $1116\pm234$~cts/s the PL signal detected from the PCW is a factor of $\sim4\times$ stronger than on the PC membrane. 
 \begin{table*}
 \begin{center}
 \begin{tabular}{|c|c|c|c|c|c|c|}
 \hline
 Position & $\tau$ (ns) & $A$ (Gcts/s$^2$) & $I$ (cts/s) & $I_{total}$ (cts/s) \\ \hline \hline
 \multirow{2}{*}{\includegraphics[height=3.7mm]{A} Bulk} & \multirow{2}{*}{$0.60\pm0.1$}  & \multirow{2}{*}{$111\pm14$} &  \multirow{2}{*}{$67\pm14$} & \multirow{2}{*}{$67\pm14$} \\[-0.5ex]
&&&& \\[-0.5ex] \hline
 \multirow{2}{*}{\includegraphics[height=3.7mm]{B} Photonic Crystal Membrane} & $0.81\pm0.1$ & $106\pm12$ & $86\pm14$ & \multirow{2}{*}{$282\pm32$} \\
                                                           & $3.92\pm0.1$ & $40\pm4$   & $196\pm18$& \\ \hline
 \multirow{2}{*}{\includegraphics[height=3.7mm]{C} Photonic Crystal Waveguide}& $0.42\pm0.1$ & $2048\pm87$& $860\pm208$& \multirow{2}{*}{$1116\pm234$} \\
   																											 & $1.74\pm0.16$& $147\pm7$  & $256\pm26$& \\ \hline
 \end{tabular}
 \end{center}
  \caption{Overview on the extracted decay times and amplitudes for \protect\includegraphics[height=3.7mm]{A} the bulk material, \protect\includegraphics[height=3.7mm]{B} the photonic crystal membrane, and \protect\includegraphics[height=3.7mm]{C} the photonic crystal waveguide.}\label{tabledecays}
 \end{table*}
This enhancement corresponds to the spatial influence of the waveguide mode. The higher photonic density of states in spatial and spectral resonance with the PCW mode reduces the radiative lifetime of the electron-hole pairs via the Purcell effect.~\cite{Purcell46} Since radiative and non-radiative recombination are competing processes in self-assembled QDs and typical non-radiative recombination rates are comparable to the radiative emission rates of QDs emitting into the photonic bandgap,~\cite{Johansen08} a reduction in the radiative lifetime will effectively increase the radiative quantum efficiency. As discussed above, the average lifetime of the QDs emitting into the photonic bandgap is lengthened to $~\sim4-5$~ns in our sample. This value is comparatively small when compared to literature where values of $\sim10-12$~ns~\cite{Kaniber07, Kaniber08a} and up to $\sim20$~ns~\cite{Lund-Hansen08, Thyrrestrup10} have been reported. We attribute this difference to a distinct non-radiative recombination rate in this specific sample, possibly due to non-ideal growth or etching conditions. Together with a smaller effective air fill factor at the position of the PCW, the suppression of non-radiative recombination due to Purcell enhanced radiative recombination can explain the increase in intensity detected from the waveguide region.
Finally, compared to the emission from QDs in the bulk material, the measured intensity of 1116$\pm234$~cts/s from the waveguide is a factor of $\sim17\times$ higher in good agreement to the emission enhancement of $\sim21\times$ extracted from Figs.~\ref{figure01} and \ref{figure02}. Furthermore, the measured lifetimes of $0.15 - 0.5$~ns are slightly reduced compared to the lifetime of QDs in the bulk material of $0.5 - 0.6$~ns, which leads to a moderate Purcell factor of $1-4$.\\

In summary, we have demonstrated a $\sim21\times$ higher out-of-plane emission intensity for quantum dots spectrally and spatially in resonance with the photonic crystal waveguide mode, as compared to quantum dots located in the unpatterned bulk material. By comparing the emission dynamics for the different photonic environments we could identify this emission enhancement as a combination of angular redistribution of emission and Purcell enhanced emission of quantum dot transitions into the waveguide mode and subsequent scattering into vertical directions. This broadband enhancement of the photoluminescence rate and intensity can be used for the construction of efficient light emitting diodes or single photon sources for optical computation applications.\\

\section*{Acknowledgements}
We gratefully acknowledge financial support of the DFG via the SFB 631, the German Excellence Initiative via NIM, the EU-FP7 via SOLID, and the BMBF via QuaHLRep project 01BQ1036. AL acknowledges support of the TUM-GS, and SF of the Alexander von Humboldt Foundation.

\section*{References}
\bibliography{Papers}
\end{document}